# Alternative Analysis Methods for Time to Event Endpoints under Non-proportional Hazards: A Comparative Analysis

Ray S. Lin[1], Ji Lin[2], Satrajit Roychoudhury[3], Keaven M. Anderson[4], Tianle Hu[5], Bo Huang[6], Larry F Leon[1], Jason J.Z. Liao[4], Rong Liu[7], Xiaodong Luo[8], Pralay Mukhopadhyay[9], Rui Qin[10], Kay Tatsuoka[11], Xuejing Wang[12], Yang Wang[13], Jian Zhu[14], Tai-Tsang Chen[11], Renee Iacona[9], Cross-Pharma Non-proportional Hazards Working Group[15]

1 Genentech/Roche, South San Francisco, CA 94080

2 Sanofi US, Cambridge, MA 02142

3 Pfizer Inc, New York, NY 10017

4 Merck Research Laboratories, North Wales, PA 19454

5 Sarepta Therapeutics, Cambridge, MA 02142

6 Pfizer Inc., Groton CT 06340

7 Celgene Co., Summit, NJ 07901

8 Sanofi US, Bridgewater, NJ, 08807

9 Astra Zeneca, Washington, DC 20004

10 Janssen Research & Development, LLC, Raritan, NJ 08869

11 Bristol-Myers Squibb, Lawrenceville, NJ 08648

12 Eli Lilly and Company, Indianapolis, IN 46285

13 Teclison Ltd., Montclair, NJ 07042

14 Takeda Pharmaceuticals, Cambridge, MA 02139

15 The Cross-Pharma NPH working group includes all the authors of this manuscript as listed above and the following members who have contributed tremendously to this work:

Prabhu Bhagavatheeswaran. Julie Cong, Margarida Geraldes, Dominik Heinzmann, Yifan Huang, Zhengrong Li, Honglu Liu, Yabing Mai, Jane Qian, Li-an Xu, Jiabu Ye, Luping Zhao

**ABSTRACT**

The log-rank test is most powerful under proportional hazards (PH). In practice, non-PH patterns are often observed in clinical trials, such as in immuno-oncology; therefore, alternative methods are needed to restore the efficiency of statistical testing. Three categories of testing methods were evaluated, including weighted log-rank tests, Kaplan-Meier curve-based tests (including weighted Kaplan-Meier and Restricted Mean Survival Time, RMST), and combination tests (including Breslow test, Lee's combo test, and MaxCombo test). Nine scenarios representing the PH and various non-PH patterns were simulated. The power, type-I error, and effect estimates of each method were compared. In general, all tests control type I error well. There is not a single most powerful test across all scenarios. In the absence of prior knowledge regarding the PH or non-PH patterns, the MaxCombo test is relatively robust across patterns. Since the treatment effect changes overtime under non-PH, the overall profile of the treatment effect may not be represented comprehensively based on a single measure. Thus, multiple measures of the treatment effect should be pre-specified as sensitivity analyses to evaluate the totality of the data.

1. **INTRODUCTION**

Time-to-event outcomes are often used as the primary endpoint for clinical trials in many disease areas. Most randomized controlled trials with a time-to-event outcome are designed and analyzed using the log-rank test and the Cox model under the assumption of proportional hazards. The log-rank p-value evaluates the statistical significance of the treatment effect, and the hazard ratio (HR) from the Cox model is used to quantify such effect. The log-rank test is most powerful and the Cox model provides unbiased HR estimates under proportional hazards (PH). However, under non-proportional hazards (non-PH), the log-rank test loses power and interpretation of the HR becomes challenging. In practice, the PH assumption is restrictive and for various reasons non-proportional hazards (non-PH) are often observed in clinical trials. In particular, patterns of delayed treatment effects have been observed recently across immuno-oncology trials. There could be multiple underlying causes for the delayed treatment effects, for example, the unique mechanism of action of the treatment, heterogeneous underlying population subgroups, and study design. Log-rank test is still statistically valid under non-PH, but it often suffers from substantial power loss. To mitigate the power loss, an increase in the sample size and/or a delay in study readout is needed, which often delays the availability of the therapy to patients with unmet medical needs. Alternative tests and estimation methods under non-PH for primary analysis may reduce false negative results whilst maintaining control of false positive rate and provide more comprehensive description of the treatment effect. They may also shorten the study duration as well as the time to bring new treatments to patients,

The most common types of non-PH, in the order of importance, are delayed treatment effects, crossing hazards, and diminishing treatment effects over time. A wide range of statistical methods for analyzing time-to-event data with different types of non-PH are discussed in the literature notably weighted log-rank tests (e.g., Fleming and Harrington 1981), weighted Kaplan-Meier tests (Pepe and Fleming 1991), restricted mean survival time comparisons (Royston and Parmar 2013), and combination tests (Breslow, et al. 1984 and Logan, et al. 2008). While there may be hypotheses about the exact nature of treatment effects at the stage of study design, we have found that such assumption is often times speculative and sometimes woefully inaccurate. This poses an additional challenge while choosing the primary analysis at the design stage of a trial with potential non-PH. Therefore, a test for the primary analysis is needed that is robust under different types of non-PH scenarios. In this paper we focus on three categories of methods as potential candidates for primary analysis.

The first category of methods includes the weighted version of the log-rank test which considers certain time periods more relevant than others. For instance, in immuno-oncology where there is a delayed treatment effect, events observed at later time points may be more precisely representing the full treatment benefit compared to the events observed at earlier time points. We have considered the Fleming–Harrington ($FH(\rho, \gamma)$) class of weighted log-rank tests. For many given underlying assumptions on treatment effects, appropriate selection of $\rho$ and $\gamma$ can provide a well-powered test by varying weights appropriately over time. The second category includes tests based on the Kaplan-Meier curve. We consider the weighted Kaplan-Meier (WKM) test and restricted mean survival time (RMST) comparisons which has gained significant attention in recent years. Finally, we consider a set of combination tests which is an adaptive procedure to select the best test from a small pre-specified set of test statistics, including multiplicity correction. In this paper we outline these three categories of test statistics and compare their operating characteristics via simulation studies.

## 2. METHODS

## 2.1. Weighted Log-rank Tests

Weighted log-rank test statistics take the form of the weighted sum of the differences of the estimated hazard functions at each observed failure time. As a result, these statistics are used to test whether the hazard difference is zero between the treatment group and the control group. In the non-PH setting, the relative differences of the two hazard functions are not constant over time, therefore a differential weighting (compared to equal weighting in the log-rank statistic) at different time points has the potential to improve the efficiency of the test statistics. In this simulation study, we are particularly interested in the Fleming-Harrington family of weighted log-rank test statistics, commonly denoted as $FH(\rho, \gamma) = S(t-)^\rho (1 - S(t-))^\gamma, \rho, \gamma \geq 0$. $FH(0,0)$ is the log-rank statistic that is most powerful under the proportional hazards assumption; when this is a diminishing effect (i.e. early separation), $FH(\rho, 0)$ with $\rho > 0$ that over weights the early events will provide higher power to detect a treatment difference compared with equal weighting; on the contrary, when delayed effect exists, $FH(0, \gamma)$ with $\gamma > 0$ that over weights the late events will be more powerful to detect the late separation; and $FH(\rho, \gamma)$ with $\rho = \gamma > 0$ will be more powerful if the biggest separation of two hazard functions occurs in the middle.. The weights in the weighted log-rank tests can be incorporated into the Cox model to provide a HR estimate of the "weighted" treatment effect (Sasieni 1993) or to provide a HR estimate of the "full" treatment effect together with a time-varying effect profile (Lin 2017). If one assumes the log-hazard ratio takes the form $\beta \Phi(t)$, where $\Phi(t)$ is a known function, then the score test for $\beta = 0$ will reduce to a weighted log-rank statistic with weight equal to $\Phi(t)$ (Lin 2017). This weight is optimal and achieves the highest testing power if the assumed hazard ratio is correct (Schoenfeld 1981).

## 2.2. Weighted Kaplan-Meier Tests

Weighted Kaplan-Meier tests take the form of the weighted sum of the differences of the Kaplan-Meier estimates (Kaplan and Meier, 1958) of survival functions. Therefore, they are valid to test whether the two underlying survival functions are the same or not. A particularly interesting weighted Kaplan-Meier test is (Pepe and Fleming, 1989, 1991) to set the weight equal to 1, resulting in the difference of two RMSTs (Uno et al. 2015, Zhao et al. 2016).

**2.3. Combination Tests**

A new set of test statistics may be derived by combining some members within a class and/or across classes. This is potentially useful in the presence of non-PH. In this comparison study, we are mainly interested in the maximum combination (MaxCombo) using the Fleming-Harrington weight family $FH(\rho, \gamma), \rho, \gamma \geq 0$ test statistics, which takes the form:

$$Z_{max} = max_{\rho, \gamma}\{Z_{FH(\rho, \gamma)}\}$$

where $Z_{FH(\rho, \gamma)}$ is the standardized Fleming-Harrington weighted log-rank statistics. In particular, we are interested in the combination of $FH(0,0), FH(0,1), FH(1,1)$ and $FH(1,0)$, which should be sensitive to PH, late-separation, middle-separation and early-separation scenarios (Fine 2007). A similar combination method incorporating only $FH(0,1)$ and $FH(1,0)$ was proposed by Lee (Lee 2007).

When the MaxCombo test is used, the treatment effect estimate is taken as the estimated HR obtained from the weighted Cox model corresponding to the weighted log-rank test with the smallest p-value.

There are other types of maximum combination tests in the literature. For example, Breslow et al. (Breslow 1984) proposed a test that is shown to be more powerful under crossing hazards compared with the log-rank test and the Peto-Prentice test (Prentice 1978); Yang et al. (Yang 2005) proposed a two-component maximum test with one component as log-rank and the other component being a customer-chosen weighted log-rank test; Yang and Prentice (Yang 2010) proposed a test that achieves symmetry between treatment and control based on their adaptively weighted log-rank test (Yang 2010); in the same paper, Yang and Prentice also proposed modified version of the maximum test and the Breslow test adding the adaptively weighted log-rank test components. However, Prentice and Yang's adaptive weighted log-rank tests are shown to have inflated Type-1 errors (Chauvel 2014 and Lin 2017), and the two-component maximum test may not be flexible enough; and therefore these methods were not included in our evaluation. We included the Breslow test because of its potential power gain under crossing hazards.

A more detailed description of methods, along with relevant references, can be found in Appendix I.

## 3. SIMULATION STUDIES

### 3.1. Simulation Study Design

In this simulation study, we used piece-wise exponential models to generate simulated data with parameters calibrated based on real trial data in order to represent common non-PH patterns, such as delayed treatment effects, diminishing treatment effects and crossing hazards, regardless of the underlying causes that can be challenging to identify. Despite the simplicity of the piece-wise exponential distribution, data simulated from these distributions mimic observed trial results quite closely suggesting that the performance of the analysis methods evaluated in this simulation study would be relevant in real studies.

All simulated trials described here equally allocated patients to an experimental arm (E) and a control arm (C). Survival data for both arms in all trials were simulated from piece-wise exponential distributions with one change point in the hazard function. The hazard change point was set as the same for the experimental arm and the control arm throughout the study. To be specific, let $\lambda_C(t)$ and $\lambda_E(t)$ denote the hazard functions, CP denote the change point, and $\lambda_{Cj}$ and $\lambda_{Ej}$ denote the hazard rates for the control arm and experimental arm respectively, where j=1 refers to the period before the change point and j=2 refers to that after the change point. The hazard functions considered in this study can be specified as $\lambda_C(t)= \lambda_{C1}*1\{0 \leq t < CP\} + \lambda_{C2}*1\{t \geq CP\}$ and $\lambda_E(t)= \lambda_{E1}*1\{0 \leq t < CP\} + \lambda_{E2}*1\{t \geq CP\}$. The piece-wise hazard ratio $HR_j = \lambda_{Ej}/\lambda_{Cj}$ was used to define the treatment effect for the jth period.

Nine scenarios were considered: seven non-PH scenarios (two *Delayed effects*, one *Diminishing effects*, and two *Crossing hazards, two Delayed effect with converging tails*), one scenario in which the PH assumption holds, and one null scenario in which there is no treatment difference. To better illustrate survival kinetics, survival functions from each scenario except for the null scenario are displayed in **Figure 1**. The parameters used to simulate data for each scenario are included in **Table 1**.

The first five non-PH scenarios have one change point, therefore, two pieces of hazard ratios that take different values. In the *Delayed effect* scenario 1, the experimental and control arms have almost identical hazards ($HR_1$=0.99) before 3 months, and the hazard decreases for the experimental arm but increases for the control arm ($HR_2$=0.478) after 3 months. The experimental and control arms in the *Delayed effect* scenario 2 also have similar hazard ($HR_1$=0.929) before 3 months, but the hazards before change point are higher than that in scenario 1, suggesting more events will occur in the first 3 months in scenario 2 compared with scenario 1. After 3 months, the hazard for the control arm remains the same while hazard for the experimental arm decreases substantially ($HR_2=0.356$).

In the *Diminishing effect* scenario, the treatment is effective (constant $HR_1=0.731$) within the first 6 months, but then the treatment effect disappears ($HR_2=0.979$) after change point.

**Figure 1. Survival Plot for Each Scenari**

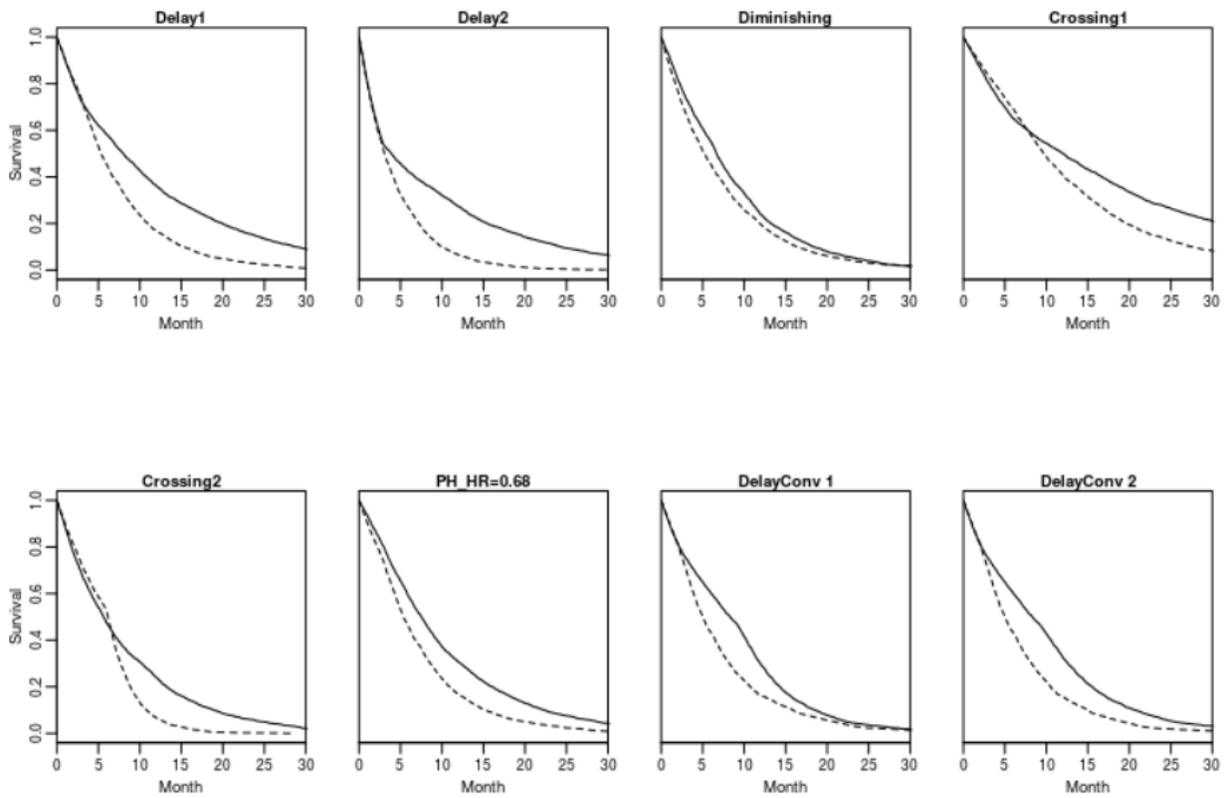

**Table 1: Parameters of Piece-wise Exponential Survival Functions for Each Scenario**

| Scenario | Single CP (months) | Time period 1 $0 \leq t < CP$ | | | Time period 2 $t \geq CP$ | | |
|---|---|---|---|---|---|---|---|
| | | $\lambda_{C1}$ | $\lambda_{E1}$ | $HR_1$ | $\lambda_{C2}$ | $\lambda_{E2}$ | $HR_2$ |
| **Delayed Treatment 1** | 3 | 0.104 | 0.103 | 0.990 | 0.161 | 0.077 | 0.478 |
| **Delayed Treatment 2** | 3 | 0.226 | 0.210 | 0.929 | 0.222 | 0.079 | 0.356 |
| **Diminishing Effect** | 6 | 0.134 | 0.098 | 0.731 | 0.140 | 0.137 | 0.979 |
| **Crossing Hazards 1** | 6 | 0.061 | 0.068 | 1.115 | 0.090 | 0.048 | 0.533 |
| **Crossing Hazards 2** | 6 | 0.108 | 0.123 | 1.139 | 0.334 | 0.120 | 0.359 |
| **Proportional Hazards** | 3 | 0.104 | 0.071 | 0.680 | 0.161 | 0.110 | 0.680 |
| **Null** | 3 | 0.104 | 0.104 | 1.000 | 0.161 | 0.161 | 1.000 |

| Scenario | 2 CPs (months) | Time period 1 $0 \leq t < CP1$ | | | Time period 2 $CP1 \leq t < CP2$ | | | Time period 3 $t \geq CP2$ | | |
|---|---|---|---|---|---|---|---|---|---|---|
| | | $\lambda_{C1}$ | $\lambda_{E1}$ | $HR_1$ | $\lambda_{C2}$ | $\lambda_{E2}$ | $HR_2$ | $\lambda_{C3}$ | $\lambda_{E3}$ | $HR_3$ |
| **DelayConv1** | 2, 7 | 0.104 | 0.103 | 0.990 | 0.161 | 0.077 | 0.478 | 0.140 | 0.168 | 1.2 |
| **DelayConv2** | 2, 7 | 0.104 | 0.103 | 0.990 | 0.161 | 0.077 | 0.478 | 0.161 | .137 | 0.85 |

NOTE: CP: change point; HR: hazard ratio; DelayConv: Delayed effect with converging tails.

*Crossing hazards* scenarios 1 and 2 represent situations in which the favorable treatment changes from the control arm to the experimental arm before and after the change point, leading to the hazard ratio changing from $HR_1 > 1$ to $HR_2 < 1$.

In the last two non-PH scenarios, *Delayed effect with converging tails* scenarios 1 and 2, there are two change points with three distinct hazard ratios, representing scenarios where there is no treatment effect in the beginning of the treatment period, then treatment benefit emerges in in the middle of the treatment period (i.e., delayed effect), and later the effect diminishes again, resulting in converging tails of survival curves.

The *Proportional hazard* scenario, where the proportional hazard assumption holds ($HR_1 = HR_2 = 0.68$), is included in the study to compare all methods where the standard log-rank test is optimal (most powerful).

The *Null* scenario ($HR_1=HR_2=1$) is included to evaluate whether each method preserves the type I error rate.

It is well-known that the total number of events plays a key role in survival analysis, and the analysis of a study is often triggered when a pre-specified number of events is reached. However, study enrollment, drop out, and sample size may also impact the analysis. To explore the impact on the testing power by the event-patient ratios (i.e., number of events divided by sample size), enrollment, and drop out, we fixed the total number of events at 210 events and considered various sample sizes, event rates, and enrolment patterns. Three sample sizes 300, 600, and 1200 (or correspondingly, three event rates 70%, 35%, and 17.5%), with three enrollment patterns were explored, resulting in a total of 9 cases within each scenario. Drop-out time is assumed to be independent of the events and follow an exponential distribution with a hazard rate of 0.014. When a total of 210 events had occurred, a data cut (i.e., administrative censoring) was applied and data were analyzed using each method.

For each of the three sample sizes, three enrollment durations were considered, namely 12, 18, and 24 months of overall enrollment duration, including a 6-month ramp-up period.

To obtain a more precise estimate of the type I error, 20,000 trial datasets were simulated for each case in the *Null* scenario. For all other scenarios, 5,000 trial datasets were simulated.

Hypothesis testing was conducted at the one-sided 2.5% significance level, and the power of each testing method was summarized. For weighted log-rank, Lee's method and MaxCombo tests, the HR estimates for the "weighted averaged" effect (Sasieni 1991) were also reported.

The simulations and analyses were conducted using the nphsim package in R (Wang et al. 2018).

## 3.2. Simulation Results

### 3.2.1. Type I Error

All nine tests under the null hypothesis of no treatment group difference control overall type I error well across the combinations of sample size and enrollment pattern. **Table 2** shows the results from the 18-month enrollment pattern and similar results were observed in the 12-month and the 24-month enrollment patterns, for which the results were included in Appendix III. Random spikes over 2.5% are mostly within simulation standard error, which is 0.1% based on 20,000 random samples. For sample sizes of 600 and 1200, the overall type I errors for the Breslow combo test tend to be much smaller, between 1.0% and 1.5%, due to the conservative nature of that test assuming asymptotic independence between component tests.

**Table 2. Overall Type I Error (%) Control**

| Sample Size | Log.Rank | FH(0,1) | FH(1,0) | FH(1,1) | RMST | WKM | Combo. Breslow | Max-Combo | Lee's |
|---|---|---|---|---|---|---|---|---|---|
| **300** | 2.590 | 2.630 | 2.520 | 2.605 | 2.545 | 2.575 | 2.505 | 2.595 | 2.565 |
| **600** | 2.585 | 2.430 | 2.770 | 2.380 | 2.590 | 2.730 | 1.210 | 2.415 | 2.445 |
| **1200** | 2.495 | 2.450 | 2.605 | 2.485 | 2.565 | 2.635 | 1.325 | 2.590 | 2.565 |

Note: RMST: restricted mean survival times; WKM: weighted Kaplan-Meier

### 3.2.2. Empirical Power

*Delayed treatment effect:* The power of the alternative tests for each simulation scenario is shown in **Figure 2** (based on the 18-month enrollment pattern; similar results were observed for the 12-month and the 24-month enrollment patterns). FH(0,1) puts more weight on late time points and therefore achieves the highest power among all the tests when there is an underlying delayed benefit. FH(1,1) puts more weights on both the middle and late time points and outperforms the log-rank test, whereas FH(1,0) puts less weights on late time points and thus performs worst among all the tests. Although not matching the performance of the FH(0,1) and FH(1,1) weighted log-rank tests, RMST generally performs better than the log-rank test while the WKM performs worse than the log-rank test. For the combination tests, the MaxCombo and Lee's tests achieve similar power to the FH(0,1) test and is consistently ranked among the best performing methods across scenarios. The Breslow-Combo test outperforms the log-rank test when the sample size is 300 (i.e., at higher event-patient ratio) yet has substantially lower power when the sample sizes are 600 and 1200 due to heavy censoring, which is consistent with the lower type-I error discussed above.

*Crossing hazards:* The power of the alternative tests for crossing hazards is similar to those of delaying treatment effect. As treatment effects are reversed and become much stronger at later time points, FH(0,1) is expected to achieve the highest power among all nine tests. The three combination tests, Breslow-Combo, MaxCombo and Lee's, are very close second to the FH(0,1), and have a clear advantage over log-rank test. FH(1,1) has a similar power as the log-rank test while RMST and WKM may have even lower power than the log-rank test, indicating that RMST and WKM tests may not necessarily perform as well under crossing hazards. The power drops significantly when the event-patient ratio decreases (i.e., when the analysis includes predominantly early events).

**Figure 2. Empirical Power of Alternative Tests for Non-Proportional Hazards for Various Sample Sizes.**

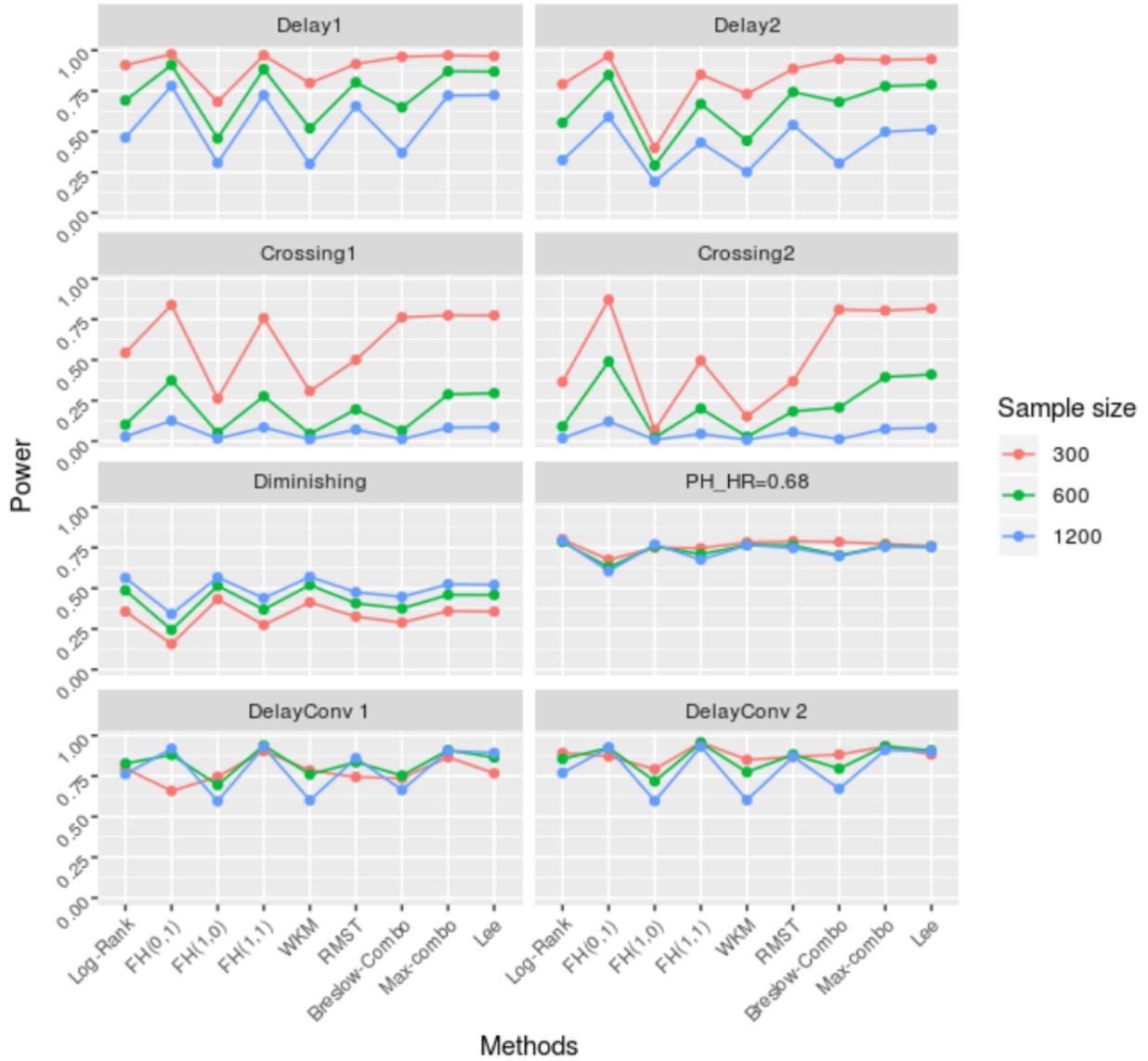

NOTE: HR: hazard ratio; PH: proportional hazard; RMST: restricted mean survival times; WKM: weighted Kaplan-Meier; DelayConv: delayed effect with converging tails.

*Diminishing treatment effect:* The power of the tests under diminishing effect size over time is compromised in the scenario we simulated. Under the diminishing effect scenario, the treatment effect is most pronounced at the earlier time points, therefore FH(1,0) is expected to achieve the highest power and FH(0,1) the lowest power. WKM has slightly lower power than FH(1,0), and is better than the log-rank test. The MaxCombo and Lee's tests have about 4% less power than the log-rank test, and slightly higher than RMST, FH(1,1) and the Breslow-Combo test.

*Proportional hazards at HR=0.68 or other:* Under the proportional hazard scenario, the log-rank test is the most powerful, however the alternative tests we considered are quite competitive, mostly within 10% power difference. The MaxCombo and Lee's test has about 3 to 4% less power compared to the log-rank test.

*Delayed effect with converging tails:* under both scenarios, FH(1,1) has the highest power as it puts more weights on the middle time points. The MaxCombo test achieves second highest power and consistently outperforms other tests across different sample sizes. Lee's and RMST tests have a similar power to the log-rank test. Furthermore, in the cases with higher event-patient ratio (e.g., sample size of 300), the power of FH(1,1) and MaxCombo remains high whereas the power of Lee's and RMST tests decreases. This is because FH(1,1) and MaxCombo tests allocate high weights on the middle time points whereas FH(1,0), FH(0,1), and Lee's tests only have the option to allocate high weights on early or late time points where the effect has not fully emerged or has started diminishing. Similarly, RMST does not down-weight late time points, and therefore its performance is impacted by the diminishing effect. In the cases with low event-patient ratio (e.g., sample size of 1200), the diminishing effect appears to have limited impact on the performance of FH(0,1), Lee's and RMST because the diminishing effect is not yet observed due to the short follow-up time.

3.2.3. Additional observations on the effects of event-patient ratio (analysis timing) and enrollment pattern

Note that in the delayed effect and crossing effect scenarios, all methods have higher power in simulated studies with 300 patients (70% event rate) compared to those with 1200 patients (17.5% event rate). This is mainly because the analysis is driven by a fixed number of events (210 events) and therefore studies with smaller sample size will have higher event-patient ratios (which includes more late events). The more mature data are able to reflect the treatment benefit after the delayed period, which increases the power. In contrast, in the diminishing effect scenario, the power increases when the sample size increases because simulated studies with larger sample size include mainly early events when the treatment effect is stronger and thus have higher power across all methods. In the PH scenario, sample size does not impact the power since the treatment effect is constant over time and thus the power mainly depends on the number of events. On the other hand, the hazard ratio is constant over time under the PH scenario, therefore the power depends only on the number of events and is similar across various event-patient ratios.

Enrollment pattern has minimal impact on the performance of the tests based on the 3 enrollment patterns simulated in this study.

3.2.4. Hazard Ratio Estimation

One way to report the treatment effect estimate when using the weighted log-rank test and the MaxCombo test is through the "weighted" HR estimated from the corresponding weighted Cox model (Sasieni 1993). The geometric means of these HR estimates are summarized in **Table 3** (based on the 18-month enrollment pattern; similar results were observed for the 12-month and the 24-month enrollment patterns). For PH scenarios (including null), the estimates of HR from the Cox model are unbiased, whereas the estimates of HR from the MaxCombo are slightly lower; 0.94-0.95 versus 1 for the null case and 0.65 versus true 0.68 for the PH case, respectively.

This slight bias (anti-conservative) is due to the model selection inherent in the MaxCombo method. Note that the model selection bias is fully addressed by the multiplicity control in hypothesis testing: the adjusted p-value procedure preserves the type-I error.

**Table 3. Geometric mean of hazard ratio estimates**

|             | Delay 1 |      | Delay 2 |      | Diminishing |      | Crossing 1 |      | Crossing 2 |      |
|-------------|---------|------|---------|------|-------------|------|------------|------|------------|------|
| Sample Size | Cox     | Max  | Cox     | Max  | Cox         | Max  | Cox        | Max  | Cox        | Max  |
| 300         | 0.63    | 0.53 | 0.68    | 0.54 | 0.8         | 0.75 | 0.75       | 0.62 | 0.8        | 0.6  |
| 600         | 0.71    | 0.58 | 0.75    | 0.61 | 0.77        | 0.73 | 0.91       | 0.76 | 0.92       | 0.72 |
| 1200        | 0.77    | 0.63 | 0.81    | 0.69 | 0.75        | 0.72 | 1          | 0.87 | 1.03       | 0.87 |
| Sample Size | PH      |      | Null    |      | DelayConv 1 |      | DelayConv 2|      |            |      |
| 300         | 0.68    | 0.65 | 1       | 0.94 | 0.68        | 0.61 | 0.64       | 0.58 |            |      |
| 600         | 0.68    | 0.65 | 1       | 0.94 | 0.67        | 0.58 | 0.66       | 0.56 |            |      |
| 1200        | 0.68    | 0.65 | 1       | 0.95 | 0.70        | 0.57 | 0.69       | 0.57 |            |      |

Note: Max: MaxCombo

For the non-PH scenarios (delayed treatment effect, diminishing treatment effect and crossing hazards), HR estimates from the MaxCombo method are consistently lower than those from the Cox model. The Cox HR estimate is an unweighted average of the treatment effect across over time whereas the MaxCombo method selects the weight function that maximizes the Z statistics (and produces the smallest HR estimate), hence reflecting an average HR that down-weights where the treatment is less effective. The slight bias due to model selection described in the PH scenarios may also contribute to the stronger effect estimates. The weighted HR can be considered an estimate that attempts to show the treatment effect focused on where the treatment is effective; this could also be done with descriptive measures such as the survival difference at different milestones or piecewise hazard ratios or piecewise hazard rates.

In the delayed treatment effect and crossing hazard scenarios, both the Cox model and MaxCombo methods have smaller HR estimates with 300 patients compared to those with 1200 patients due to the additional data maturity described above. In contrast, in the diminishing effect scenario, HR estimates decrease as the sample size increases (and event-patient ratio decreases).

In the PH scenario (including null), sample size does not impact the HR estimate since the treatment effect is constant over time.

## 4. REAL DATA EXAMPLES

To explore alternative tests in real clinical studies, Bristol-Myers Squibb (BMS) and AstraZeneca contributed survival datasets from two completed oncology clinical studies. The KM plots and statistical results are presented in the Appendix II. We compared the time-to-event endpoints using the datasets reconstructed (Guyot et al. 2012) based on the original publication between the two treatment arms using the weighted log-rank test with FH(0,1), FH(1,0), FH(1,1) weights, the max-combo test (with set of weights of (0,0), (0,1), (1,0) and (1,1)) and the difference in RMST. Given the non-PH pattern observed, piece-wise HRs were also estimated with the underlying change point that was selected post hoc based on the KM curves. These methods were retrospectively applied in order to contrast with the results using log-rank tests and standard Cox HR estimates in the original publication. Note that results reported in the original publications were based on stratified analyses whereas our results were based on unstratified analyses. Minor differences were observed, but they will not affect the interpretation of the comparisons presented below.

**4.1. Case study 1**: Ipilimumab 10 mg/kg versus ipilimumab 3 mg/kg in patients with unresectable or metastatic melanoma

A randomized phase 3 trial was conducted to study the 10 mg/kg dose versus the 3 mg/kg dose in patients with untreated or previously treated unresectable stage III or IV melanoma with a primary endpoint of OS (Ascierto et al 2017). Of the 727 patients who underwent randomization, 365 (364 treated) were assigned to ipilimumab 10 mg/kg and 362 (all treated) were assigned to ipilimumab 3 mg/kg. Median OS was 15.7 months (95% CI 11.6-17.8) for ipilimumab 10 mg/kg compared with 11.5 months (95% CI 9.9-13.3) for ipilimumab 3 mg/kg (HR 0.84, 95% CI 0.70-0.99; p=0.04). The p-value was based on a stratified log-rank test and HR and associated 95% CI estimated using a stratified Cox model.

Non-PH was suspected based on visual inspection of the OS curves, which overlapped before 9 months and start to separate afterward, representing a typical "delayed effect" scenario. A Schoenfeld residual plot was generated (**Figure 3**), which showed a potential non-random pattern over time; however, the G-T p-value (Grambsch and Therneau 1994) was not significant (p=0.142). Note that this is consistent with prior literature that shows the G-T test is not a powerful test and may fail to declare statistical significance even though PH assumption is clearly violated (Lin 2017). **Table 4** contains the statistical results of the six tests along with the corresponding effect estimates and 95% CI of OS with a survival follow up of 2 years. All p-values are based on an unstratified analysis. A treatment effect delay of approximately 9 months was observed by visual inspection, so piece-wise Cox HRs with a change point at 9 months were estimated.

**Figure 3. Case Study 1 Schoenfeld Residual Plot for OS**

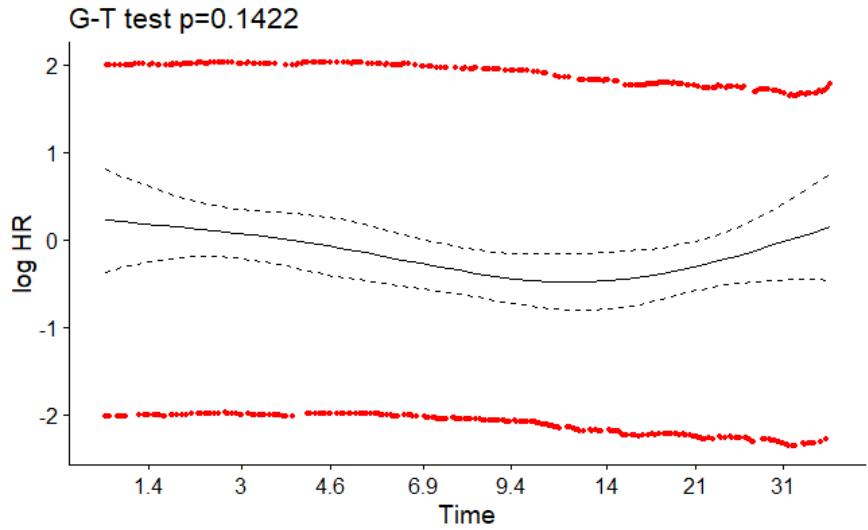

**Table 4. Case Study 1 Statistical Results of Various Methods for Overall Survival**

| Method | Two-sided p-value | Effect Size estimate | 95% CI |
|---|---|---|---|
| Log-rank test | 0.0587 | 0.850 | 0.718-1.006 |
| WLRT (FH 0,1) | 0.0174 | 0.790 | 0.649-0.960 |
| WLRT (FH 1,0) | 0.1939 | 0.888 | 0.743-1.062 |
| WLRT (FH 1,1) | 0.0095 | 0.788 | 0.658-0.944 |
| MaxCombo (selected wt = [0,1]) | 0.0208 | 0.788 | 0.643-0.967 |
| RMST difference (months) | 0.0176 | 2.683 | 0.187-5.180 |
| Piece-wise (months) Cox HR | | | |
| (0, 9) | | 0.957 | 0.762-1.201 |
| [9, inf) | | 0.737 | 0.573-0.947 |

Note: CI: confidence interval; RMST: restricted mean survival times; WLRT: weighted log-rank test; wt: weight; HR: hazard ratio

With the exception of the unstratified log-rank and FH(1,0) test, all test results were statistically significant. Note the stratified log-rank test was statistically significant (p-value 0.04). The FH(0,1) test, MaxCombo, and RMST outperformed the log-rank test. It is not surprising that the FH(1,0) test did not perform well, as it puts more weight in the early events when survival curves for the two arms overlap. It is also worth noting that the max-combo test (selected weight = [0,1]) had a slightly larger p-value than the FH(0,1), due to the penalty for the multiplicity adjustment due to inclusion of all four tests.

**4.2. Case study 2**: Gefitinib versus chemotherapy (paclitaxel/carboplatin) in first line non-small cell lung cancer

The IPASS study (Mok et al 2009) of gefitinib versus chemotherapy (paclitaxel/carboplatin) in first-line non-small cell lung cancer was a phase 3 open label trial, where patients were equally randomized to gefitinib (609 patients) or chemotherapy (608 patients). The primary endpoint was PFS evaluated in all randomized patients. OS in all randomized patients was a key secondary endpoint.

The study required 944 PFS events to have 80% power to demonstrate a noninferiority (NI) margin of 1.2 if the treatments were truly equal, with a two-sided 5% probability of incorrectly concluding NI. If NI was demonstrated, testing for superiority was conducted and the treatment was declared superior if the upper bound of the 95% CI for HR was below 1 (equivalently, the 2-sided p-value was less than 0.05).

The study demonstrated a statistically significant PFS improvement in favor of gefitinib. However, interestingly the PFS initially favored the chemotherapy arm, with the curves crossing at around the end of the 6th month in favor of gefitinib. The OS results also showed similar features of crossing OS curves, although the overall treatment effect was more modest and did not reach statistical significance at the time of analysis.

A Schoenfeld residual plot for PFS was generated (**Figure 4 a**), which showed a non-random pattern over time (G-T p-value < 0.001) formally confirming initial observations.
**Table 5** shows the results of PFS analyses from these six tests along with the corresponding treatment effect size estimates and 95% CIs. Given the PFS curves were crossing at approximately 6 months by visual inspection, HRs were also estimated separately with a change point at 6 months. Similar observations as in Case Study 1 were made in comparison of the different tests and HR estimation for PFS.

Similar analyses, as in PFS, were conducted for OS. Again, based on visual inspection and the Schoenfeld residual plot (**Figure 4 b**), data were indicative of a lack of PH, even though the G-T test was not significant (p=0.67).

**Table 5** shows the results of OS analyses from these six tests along with the corresponding treatment effect size estimates. The OS results from the unstratified log-rank test were not statistically significant. Note that the original trial reported a HR of 0.91 (95% CI, 0.76–1.10). Interestingly, only two of the six tests were statistically significant in this case: the FH(1,1) and the MaxCombo (selected weight = [1,1]). Looking at the OS curve, it makes intuitive sense why the FH(1,1) may be statistically significant, given it puts more emphasis on events occurring in the middle part of the curve versus those happening early or late. This also demonstrates the agility of the MaxCombo test to identify different patterns of non-PH without knowing a priori which one will actually be observed. Due to the multiplicity adjustment, the p-value from the MaxCombo test is again slightly larger than the FH(1,1). Nonetheless, both tests are significant. Given that the OS curves were crossing at approximately 8 months, and most patients were censored beyond 35 months to better understand how the treatment effect evolved over time, we computed HRs with change points at 8 and 35 months.

These results are suggestive of a statistically significant difference in PFS (from all tests except FH(1,0)) and OS (if being tested using the FH(1,1) or the max-combo) and a potential advantage of the MaxCombo when the underlying non-PH pattern is unknown.

**Figure 4. Case Study 2 Schoenfeld Residual Plot for PFS (a) and OS (b)**

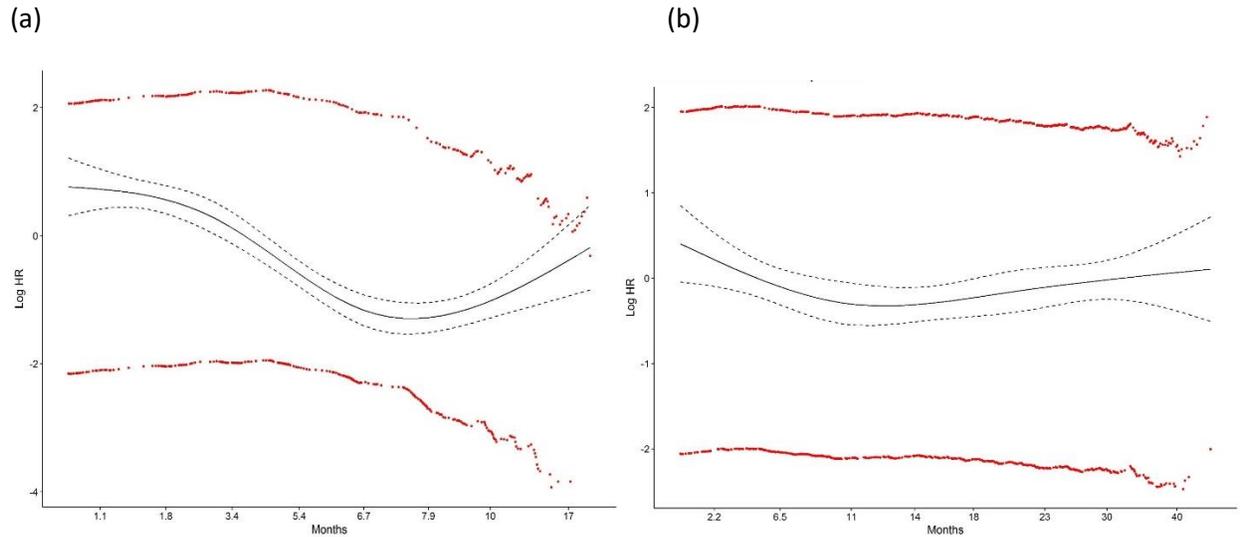

**Table 5. Case Study 2 (IPASS) Statistical Results of Various Methods for Progression-Free Survival and Overall Survival**

| Endpoint | Method* | Two-sided p-value | Effect Size estimate | 95% CI |
| --- | --- | --- | --- | --- |
| PFS | Log-rank test | <0.001 | 0.73 | 0.640 - 0.832 |
|  | WLRT (FH 0,1) | <0.001 | 0.481 | 0.410 - 0.564 |
|  | WLRT (FH 1,0) | 0.345 | 0.972 | 0.845 - 1.117 |
|  | WLRT (FH 1,1) | <0.001 | 0.596 | 0.517 - 0.687 |
|  | MaxCombo (selected wt = [0,1]) | 0.00034 | 0.481 | 0.400 - 0.578 |
|  | RMST difference (months) | p<0.001 | 1.407 | 0.785 - 2.029 |
|  | Piece-wise (months) Cox HR |  |  |  |
|  | (0, 6) |  | 1.115 | 0.948 – 1.310 |
|  | [6, inf) |  | 0.343 | 0.276 – 0.428 |
| OS | Log-rank test | 0.052 | 0.90 | 0.793 - 1.022 |
|  | WLRT (FH 0,1) | 0.053 | 0.886 | 0.764 - 1.026 |
|  | WLRT (FH 1,0) | 0.092 | 0.911 | 0.794 - 1.045 |
|  | WLRT (FH 1,1) | 0.009 | 0.850 | 0.743 - 0.973 |
|  | MaxCombo (selected wt = [0,1]) | 0.017 | 0.850 | 0.728 - 0.994 |
|  | RMST difference (months) | 0.085 | 1.201 | 0.517 - 2.918 |
|  | Piece-wise (months) Cox HR |  |  |  |
|  | (0, 8) |  | 0.99 | 0.77 – 1.28 |
|  | [8, 35) |  | 0.85 | 0.73 – 0.99 |
|  | [35, inf) |  | 1.283 | 0.687 - 2.394 |

Note: CI: confidence interval; RMST: restricted mean survival times; WLRT: weighted log-rank test; wt: weight; HR: hazard ratio

* 67 patients (19 in Gefitinib arm; 48 in chemotherapy arm) were excluded from the analysis as patients may have withdrawn consent

## 5. DISCUSSION AND RECOMMENDATION

Non-PH has been observed in immuno-oncology, for example, the delayed separation of the survival curves or even crossed survival curves. In these cases, the standard log-rank tests lose power substantially. Simulations were conducted to evaluate nine alternative tests under various scenarios for non-PH. All the tests control overall type-I error well across the combinations of sample size and enrollment pattern, with the Breslow-combo test tending to be more conservative when event rate is relatively small. None of the alternative tests were uniformly most powerful across all non-PH scenarios. Depending on the nature of the non-PH, certain tests are more robust than others under model misspecification. Particularly, without the knowledge of the non-PH pattern in advance, the MaxCombo test is robust and agnostic to various patterns of non-PH and increases the power of the test by adaptively selecting the weight function based on the observed data with control of multiplicity. It provides a strong advantage under delayed effect or crossing hazards (which are quite commonly observed in immuno-oncology), while providing acceptable power under diminishing effect and PH (3-4% loss of power) compared to the log-rank test.

The selection of a test for primary analysis should be clearly pre-specified and guided by prior knowledge of the treatment (for example, there is likely delayed effect due to the mechanism of action) and general clinical settings (for example, there is likely diminishing effect due to effective subsequent therapies that could confound long-term survival). If there is limited prior knowledge regarding the nature of the non-PH at the study design stage, a combination test (such as the MaxCombo test) could be a good alternative against the risk of losing power if the model assumption is severely violated. When using a combination test, the set of weight functions to be included in the test should also be pre-specified based on prior knowledge and clinical relevance.

Analogous to the Fleming-Harrington weight family shown in our simulations, the weights could also be pre-specified according to the quantiles of the pooled Kaplan-Meier curve [e.g., 25%, 50%, and 75% percentiles] or at specific time points (e.g., 3, 6, 12 months).

While the weighted log rank tests and the combination test can be a very useful tool when analyzing data under non-PH, we do have to be cautious that a statistically significant result may not always imply clinically meaningful improvement. For example, the MaxCombo one-sided test there may suffer from slight type-I error inflation in the scenarios where the treatment arm starts with a detrimental effect and later turns into a beneficial effect (i.e., the hazard ratio changes from greater than one to less than one) yet the survival curves do not cross (i.e., the survival is consistently lower throughout the study duration in the treatment arm compared to the control arm) (Roychoudhury, Anderson and Mukhopadhyay 2018). Therefore, once a treatment difference is shown to be statistically significant, a thorough evaluation of the treatment effect still needs to be conducted based on the totality of evidence.

An important question under non-PH is: how do we accurately estimate and report treatment effect that is changing over time? Summarizing the effect across time based on the traditional single HR estimate (which average the treatment effect over time) may potentially be misleading because the treatment does not have a constant effect throughout all time points and does not benefit all patients equally. For instance, in the Case Study 2 (IPASS) example, reporting an overall HR of 0.73 could be misleading in the context of crossing PFS curves, as clearly the benefit is much more substantial at later time points and among those patients who did not have early progressive diseases. Based on the observed data, the piece-wise constant HRs could be used to describe the change in treatment effect over time. In the real world examples, the change points of the HRs were chosen subjectively based on review of the observed Kaplan-Meier curves. However, in practice, we recommend that these change points be pre-specified in the study analysis plan based on prior knowledge about the treatment or based on clinical relevance. The PFS or OS rates at pre-specified time-points (e.g., 6, 12 months) have simple clinical interpretation and can help describing the treatment effect through multiple time points on the Kaplan-Meier curves, which can be pre-specified based on expected study duration and clinical relevance. The difference in RMST provides a different perspective in terms of quantifying the benefit and is especially appealing since it does not assume PH. Similar to piece-wise HRs, RMST estimates can be evaluated in an ad-hoc fashion as a function of survival time to profile the characteristics of non-PH nature (Zhao et al 2016). We recommend reporting these multiple measures in order to reflect the totality of the data and to convey to clinicians and patients a comprehensive view of the treatment effect for clinical decision making.

In addition, for studies designed for registration purposes, it is also important to communicate with health authorities in advance to align the statistical view on the potential non-PH and attain regulatory agreement on alternative tests. The current regulatory standard of binary decision making for declaring a study to be positive or negative, based on a single p-value (from the log-rank test) and estimating the treatment benefit using a single summary measure (i.e., HR from the Cox model) can be problematic, when non-PH is observed. This is because in such cases the benefit is clearly non-uniform among patients. If the initial test fails to detect statistical significance, further investigation will generally be considered exploratory. In this case, an experimental molecule with still a substantial benefit for many patients will not be able to receive marketing authorization. Therefore, the need for a more powerful test, such as the MaxCombo when non-PH is expected, can be critical in establishing this initial statistical difference between the two arms. Once that difference is established, and the study is declared positive, further investigation to optimize benefit-risk is possible. Similarly, when PH is violated, a single measure such as the HR may not be adequate in describing treatment benefit and use of additional measures such as piecewise HR, milestone survival and difference in RMST can be very useful in interpreting the trial results. It is recommended to develop a comprehensive analysis plan that defines the primary test, such as MaxCombo test, and the additional sensitivity analyses to evaluate the totality of the data based on alternative tests and summary statistics as well as standard analysis methods (e.g., log-rank test and Cox model). Such analysis plan could enable better characterization of the treatments.

**Acknowledgements**

This manuscript was prepared as a summary of the presentations given by the authors and the valuable comments and questions received at the 2018 Duke-Margolis Public Workshop: Oncology Clinical Trials in the Presence of Non-Proportional Hazards (https://healthpolicy.duke.edu/events/public-workshop-oncology-clinical-trials-presence-non-proportional-hazards). The authors, thereby, sincerely acknowledge the tremendous contribution of the attendees, the sponsors, the organizing committee and the speakers of the workshop.

# APPENDICES

## Appendix I Detailed Description of Methods

Detailed description of methods summarized separately in a PDF file to be inserted here (currently inserted after Appendix III for review).

**Appendix II. KM Plots and Statistical Analysis Results in Case Studies**

**Figure A.1 Case Study 1 KM Plot and Statistical Analysis Results**

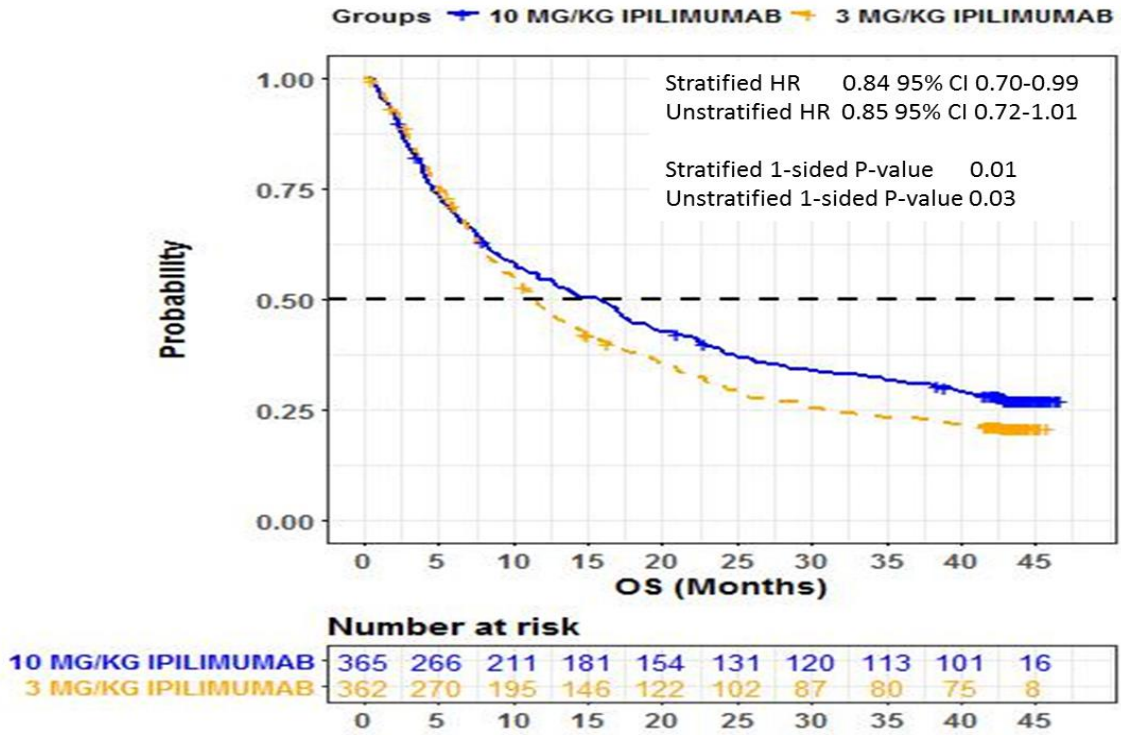

**Figure A.2. Case Study 2 (IPASS) Progression-free survival of Gefitinib Versus Chemotherapy in the Intent-to-treat Population.**

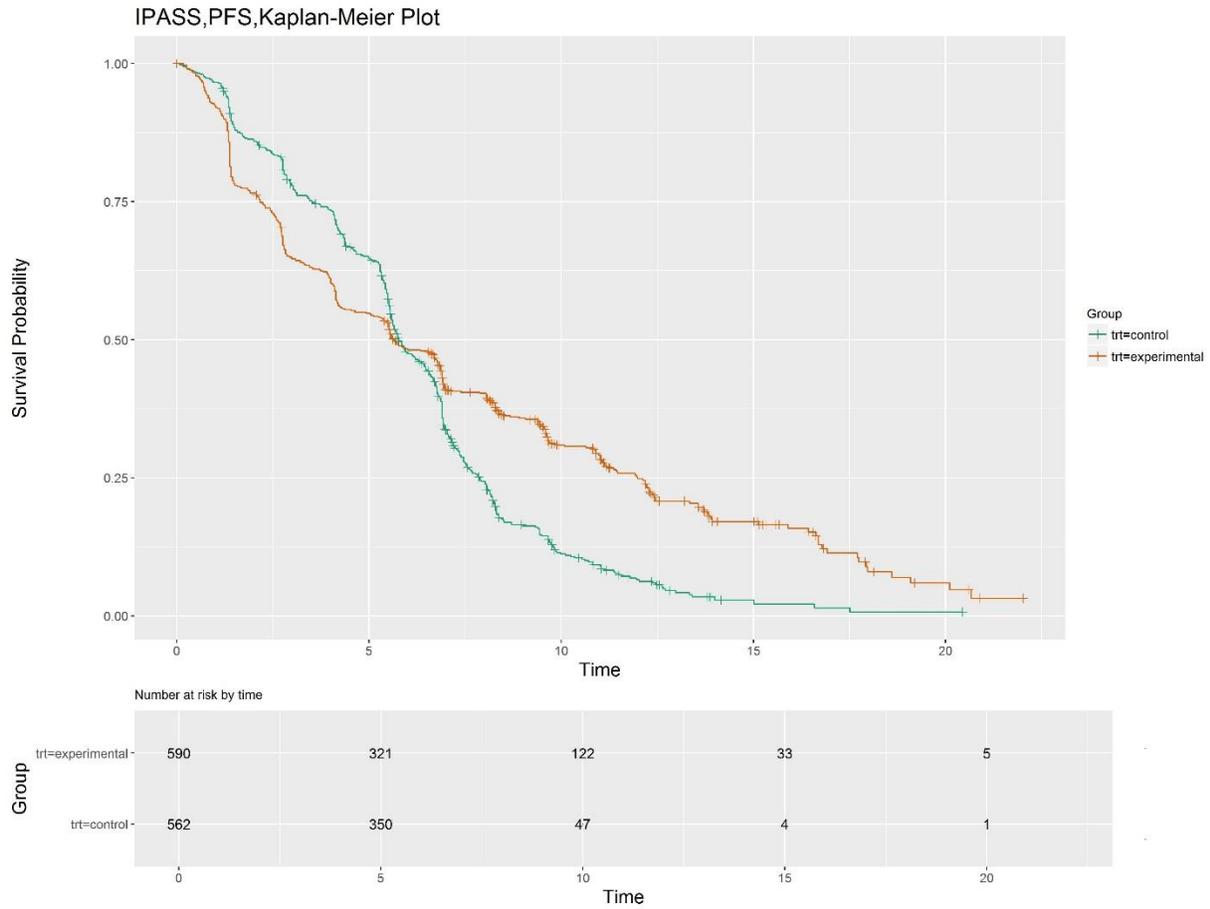

Note: 67 patients (19 in Gefitinib; 48 in Pac/Carbo) removed from analysis during patient de-identification process

**Figure A.3. Case Study 2 (IPASS) Overall Survival of Gefitinib Versus Chemotherapy in the Intent-to-treat Population**

CI: confidence interval; HR: hazard ratio

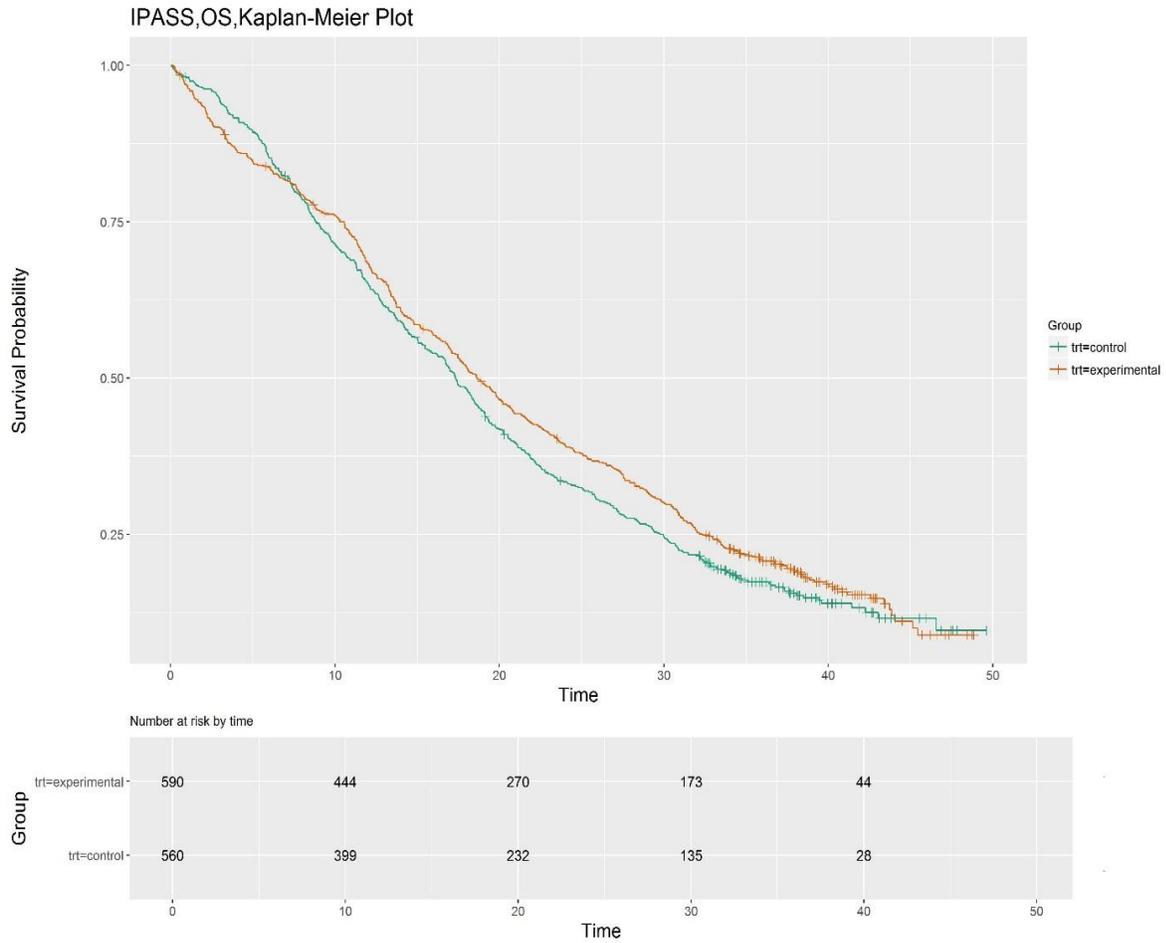

Note: 67 patients (19 in Gefitinib; 48 in Pac/Carbo) removed from analysis during patient de-identification process

**Appendix III. Overall Type I Error (%) Control in the 12-month and the 24-month enrollment patterns**

12-month enrollment pattern

| Sample Size | Log.Rank | FH(0,1) | FH(1,0) | FH(1,1) | RMST | WKM | Combo.Breslow | MaxCombo | Lee's |
|---|---|---|---|---|---|---|---|---|---|
| **300** | 2.48 | 2.49 | 2.475 | 2.48 | 2.575 | 2.5 | 2.145 | 2.545 | 2.48 |
| **600** | 2.38 | 2.385 | 2.29 | 2.415 | 2.35 | 2.3 | 1.045 | 2.3 | 2.265 |
| **1200** | 2.84 | 2.615 | 2.69 | 2.53 | 2.71 | 2.66 | 1.445 | 2.665 | 2.645 |

24-month enrollment pattern

| Sample Size | Log.Rank | FH(0,1) | FH(1,0) | FH(1,1) | RMST | WKM | Combo.Breslow | MaxCombo | Lee's |
|---|---|---|---|---|---|---|---|---|---|
| **300** | 2.465 | 2.605 | 2.445 | 2.425 | 2.275 | 2.475 | 2.595 | 2.43 | 2.43 |
| **600** | 2.295 | 2.5 | 2.46 | 2.44 | 2.4 | 2.495 | 1.385 | 2.46 | 2.445 |
| **1200** | 2.505 | 2.59 | 2.59 | 2.45 | 2.635 | 2.645 | 1.26 | 2.615 | 2.59 |

Note: RMST: restricted mean survival times; WKM: weighted Kaplan-Meier

**(Appendix I Detailed Description of Methods starting from next page)**